\documentclass[showpacs, pra,onecolumn,preprintnumbers ,amsmath, amssymb, superscriptaddress, aps]{revtex4-2}
\usepackage{color}
\usepackage{amsmath,amssymb}
\usepackage{pifont}
\usepackage{amssymb}  
\usepackage{bbold}
\usepackage{float}
\usepackage{subfloat}
\usepackage[squaren]{SIunits}

\usepackage[caption=false]{subfig}
\usepackage{tikz}
\usepackage{makecell}
\usepackage{subfig}

\usepackage{pifont}   
\usepackage{graphicx} 
\graphicspath{{Figuress/}}
\usepackage{dcolumn}  
\usepackage{bm}       
\usepackage{multirow} 
\usepackage{placeins}
\usepackage[colorlinks]{hyperref}
\usepackage{mathtools}
\usepackage{appendix}
\usepackage{amsmath}
\usepackage{appendix}

\usepackage{fancyhdr}
 \usepackage{orcidlink}
\begin{document}

  \title{\textbf{Assessing the entanglement of three coupled harmonic oscillators}}
  \author{Ayoub Ghaba}
       \email{ayoub.ghaba-etu@etu.univh2c.ma}
 \affiliation{Laboratory of Mathematics and Physical Sciences Applied to Engineering Sciences, Faculty of Sciences and Techniques, Hassan II University, Mohammedia, Morocco}
  \author{Radouan Hab Arrih}
        \email{habarrih46@gmail.com}
 \affiliation{Laboratory of R\&D in Engineering Sciences, Faculty of Sciences and Techniques Al-Hoceima,
Abdelmalek Essaadi University, Tetouan, Morocco.}
\author{Elhoussine Atmani}
\affiliation{Laboratory of Mathematics and Physical Sciences Applied to Engineering Sciences, Faculty of Sciences and Techniques, Hassan II University, Mohammedia, Morocco}
 
\author{Abderrahim El Allati}
\affiliation{Laboratory of R\&D in Engineering Sciences, Faculty of Sciences and Techniques Al-Hoceima,
Abdelmalek Essaadi University, Tetouan, Morocco.}
 \author{Abdallah Slaoui \orcidlink{0000-0002-5284-3240}}\email{abdallah.slaoui@um5s.net.ma}\affiliation{LPHE-MS, Faculty of Sciences, Mohammed V University in Rabat, Rabat, Morocco}\affiliation{CPM, Faculty of Sciences, Mohammed V University in Rabat, Rabat, Morocco.} 
  \date{\today}

  	\maketitle
 
 	\begin{center}
 	\textbf{Abstract}
 \end{center}
Quantum entanglement serves as a key phenomenon in understanding correlations in many-body systems, but analytical results remain scarce for coupled three-body oscillators. In this work, we address this gap by introducing a geometrical diagonalization approach that constrains Euler angles, thereby reducing the degrees of freedom in the entanglement analysis. It consists of deriving analytical expressions for linear entropy and purity under the bipartitions $(x|yz)$, $(y|xz)$, and $(xy|z)$ using the Wigner function framework. Our results indicate that excitations in any oscillator basically enhance the redistribution of correlations across the system. The mixing angle $\theta$ governs entanglement intensity, ranging from separability to maximal correlation. Moreover, we reveal the symmetry relations, notably $S_{Ly}[(n,m,l),\theta]=S_{Lz}[(n,m,l),-\theta]$ and an intrinsic symmetry within $(x|yz)$. Hence, we clarify how excitation levels and mixing angles create and enhance entanglement in the three coupled harmonic oscillators.\\


\section{Introduction}
Quantum entanglement is one of the most fascinating features of quantum mechanics, even though it was originally introduced to challenge some of the foundational aspects of the theory. In particular, Einstein, Podolsky, and Rosen highlighted that entangled wave functions suggest the presence of hidden variables, which revealed incompleteness in quantum mechanics \cite{1,2}. Subsequently, the entanglement phenomenon has been identified as a fundamental aspect of quantum theory \cite{3,Slaoui2023,Tarif2025}, which is essential for interpreting and validating its predictions \cite{4}. Recently, entanglement has been recognized as a crucial resource in quantum information science, enabling protocols that surpass classical communication and computational capabilities \cite{5,46,47,48,49}. In fact, the entanglement phenomenon is used to design experimental setups that can generate a robust entanglement degree, which preserves it against decoherence and enables optimal control \cite{habb1,habb2,habb3,habb4,habb5,habb6,habb7,habb8,habb9,allati1}. Additionally, quantum entanglement in continuous-variable systems highlighted a central and widely explored topic in quantum information science \cite{adesso1,Naimy2025,adesso2,adesso3,adesso4,adesso5,adesso6,adesso7,allati2,allati3}. Particularly, the entanglement using Gaussian states has been studied in systems of two and three coupled oscillators under the influence of intrinsic decoherence \cite{hab1,hab2,Alaoui2024}, extrinsic decoherence \cite{hab3,hab4,hab5,hab6} and in ideal scenarios without decoherence \cite{park1,park2,intro8}. However, quantum entanglement by means of non-Gaussian states of two coupled oscillator systems has also been explored \cite{Maka1,Maka2, Maka3,RAD1}. In fact, it is shown that the entanglement is significantly enhanced in highly excited states. This highlights the fact that quantum entanglement in non-Gaussian states can play a more prominent role than in their Gaussian counterparts \cite{Ngo,Ngo1}. Interestingly enough, the coupled harmonic oscillators provide a simple yet versatile physical platform that can be used to realize both Gaussian and non-Gaussian states.\\

Coupled harmonic oscillator systems constitute an omnipresent physical platform where many systems can be effectively modeled in the vicinity of equilibrium by coupled harmonic oscillators. Indeed, quantum oscillators play a central role in understanding fundamental phenomena such as superconductivity \cite{intro3}, quantum optics \cite{intro4}, biophysics \cite{intro9, intro10, intro11}, molecular chemistry \cite{intro12, intro13} and quantum chemistry \cite{intro14, intro15}. Importantly, Makarov investigated quantum entanglement in two coupled oscillators using the Schmidt decomposition of the quantum state \cite{Maka1, Maka2}. By introducing suitable physical assumptions, he elegantly reduced the problem of entanglement to a dependence on the mixing angle:
  \[
  \theta = \tfrac{1}{2}\tan^{-1}\!\left(\frac{2\epsilon}{\omega_1^{2}-\omega_2^{2}}\right),
  \]

where $\omega_i$ denotes the frequency of oscillator $i$, and $\epsilon$ represents the coupling strength between the oscillators. Motivated by this remarkable work, we are naturally led to ask whether a similar approach can be extended to a more complex system, three coupled harmonic oscillators. Such a system presents significantly greater challenges, both in the diagonalization procedure and in the characterization of entanglement between different partitions \cite{class}. By exploiting an appropriate mathematical identity involving the eigenvalues and eigenvectors of the system one can achieve the emergence of three distinct mixing angles $\theta,\phi,\varphi$ in terms of physical parameters, namely the couplings and frequencies \cite{intro8}. Therefore, characterizing entanglement by means of these three interdependent mixing angles turns out to be extremely complex. In this work, we introduce a new approach, geometric diagonalization, which effectively reduces the description of entanglement to a dependence on a single mixing angle, namely $\theta$. Furthermore, we will use analytical and numerical techniques in the phase space framework to evaluate the entanglement of each bipartition, rather than relying on the traditional Schmidt decomposition. Our analysis of quantum entanglement in this complex system reveals that entanglement is very sensitive to the mixing angle, $\theta$, as well as to the quantum numbers $n$, $m$, and $l$. \\

In addition, our work provides a comprehensive insight into the structure of quantum entanglement across different bipartitions, highlighting the intricate distribution of correlations within the system. This detailed understanding demonstrates that the system can serve as a highly versatile and robust resource for generating and controlling entanglement. Consequently, our physical platform can offer significant potential for implementation in various quantum technologies; including quantum communication and quantum sensing protocols. Moreover, the tunability of the system parameters and the controllable mixing angles enable flexible scenarios for encoding, processing, and manipulating quantum information. Ultimately, these features allow us to consider the coupled harmonic oscillator networks as promising candidates for the realization of scalable quantum computing architectures.\\

The remainder of this paper is organized as follows. In Sec.~\ref{sec2}, we introduce a system of three coupled harmonic oscillators and outline the geometrical diagonalization procedure. Sec.~\ref{sec3} includes the quantification of quantum entanglement using the Wigner function and linear entropy. In Sec.~\ref{sec4}, we report our results accompanied by a detailed physical discussion. Finally, Sec.\ref{sec5} provides the concluding remarks.

\section{Hamiltonian and Geometrical Diagonalization \label{sec2}  }

Let's consider a tripartite system of coupled harmonic oscillators, namely $x$, $y$, and $z$. Each oscillator possesses an intrinsic angular frequency, namely $\omega_x$, $\omega_y$, and $\omega_z$, respectively. The quadratic describes both the kinetic and potential energy contributions of the three coupled harmonic oscillators. Hence, the Hamiltonian of this system is given as \cite{hab1, park1}
\begin{equation}
\mathcal{H} =\frac{1}{2}\sum _{a=x,y,z}\left( \hat{p}_a^2+\omega_a^2 \hat{a}^2 \right)+ J_{xy} \hat{x} \hat{y} + J_{xz} \hat{x} \hat{z} + J_{yz} \hat{y} \hat{z}, \label{eq1}
\end{equation}
with $\hbar = m = 1$ \cite{1st_chapter1}, and  the operators $\hat{x}$, $\hat{y}$, and $\hat{z}$ represent the position operators of the oscillators, while their conjugate momenta are defined as $\hat{p}_a = -i\,\frac{\partial}{\partial a}$, satisfying the canonical commutation relations $[\hat{a}, \hat{p}_a] = i$, where $i$ denotes the imaginary unit. The parameters $\omega_a$ correspond to the angular frequencies of the oscillators, and $J_{ab}$ quantify the coupling strengths between oscillators $a$ and $b$. To investigate the quantum entanglement properties of the system, the Hamiltonian must be diagonalized, which can be achieved through a suitable rotation parameterized by the Euler angles $(\theta, \varphi, \Phi)$
\cite{1st_chapter2}. Indeed, we harness the following orthogonal rotation matrix \cite{intro8}
\begin{equation}
    \mathbb{R}(\varphi, \Phi, \theta) = e^{i\varphi \hat{L}_1} e^{i\Phi \hat{L}_2} e^{i\theta \hat{L}_3},
\end{equation}
Here, the generators $\hat{L}_k = (\hat{q}_i \hat{p}_j - \hat{q}_j \hat{p}_i)$ satisfy the closed algebra 
$
[\hat{L}_i, \hat{L}_j] = i\,\epsilon_{ijk}\,\hat{L}_k,
$, where $\hat{q}_i$ and $\hat{p}_i$ denote the canonical position and momentum operators, respectively, $\epsilon_{ijk}$ is the Levi-Civita antisymmetric tensor, and $i$ represents the imaginary unit. Besides, the rotation matrix can be explicitly expressed as:

\begin{equation}
    \mathbb{R}(\varphi, \Phi, \theta) = \left( \begin{array}{ccc}
C_\theta C_\Phi & S_\Phi & S_\theta C_\Phi \\
- C_\varphi S_\Phi C_\theta - S_\varphi S_\theta & C_\varphi C_\Phi & - C_\varphi S_\Phi S_\theta + S_\varphi C_\theta \\
S_\varphi S_\Phi C_\theta - C_\varphi S_\theta & - S_\varphi C_\Phi & S_\varphi S_\Phi S_\theta + C_\varphi C_\theta
\end{array} \right),\label{eq5}
\end{equation}
where we have set $
(S_\eta, C_\eta) = (\sin \eta, \cos \eta)$, with $\eta \in \{ \varphi, \Phi, \theta \}$.\par

This work aims to address the following questions: How can we devise a diagonalization scheme that imposes additional constraints among the Euler angles? Based on this, how can we reduce the corresponding number of these angles to simplify the analysis of quantum entanglement within the system?

\begin{figure}[H]
    \centering
    \includegraphics[width=0.6\linewidth]{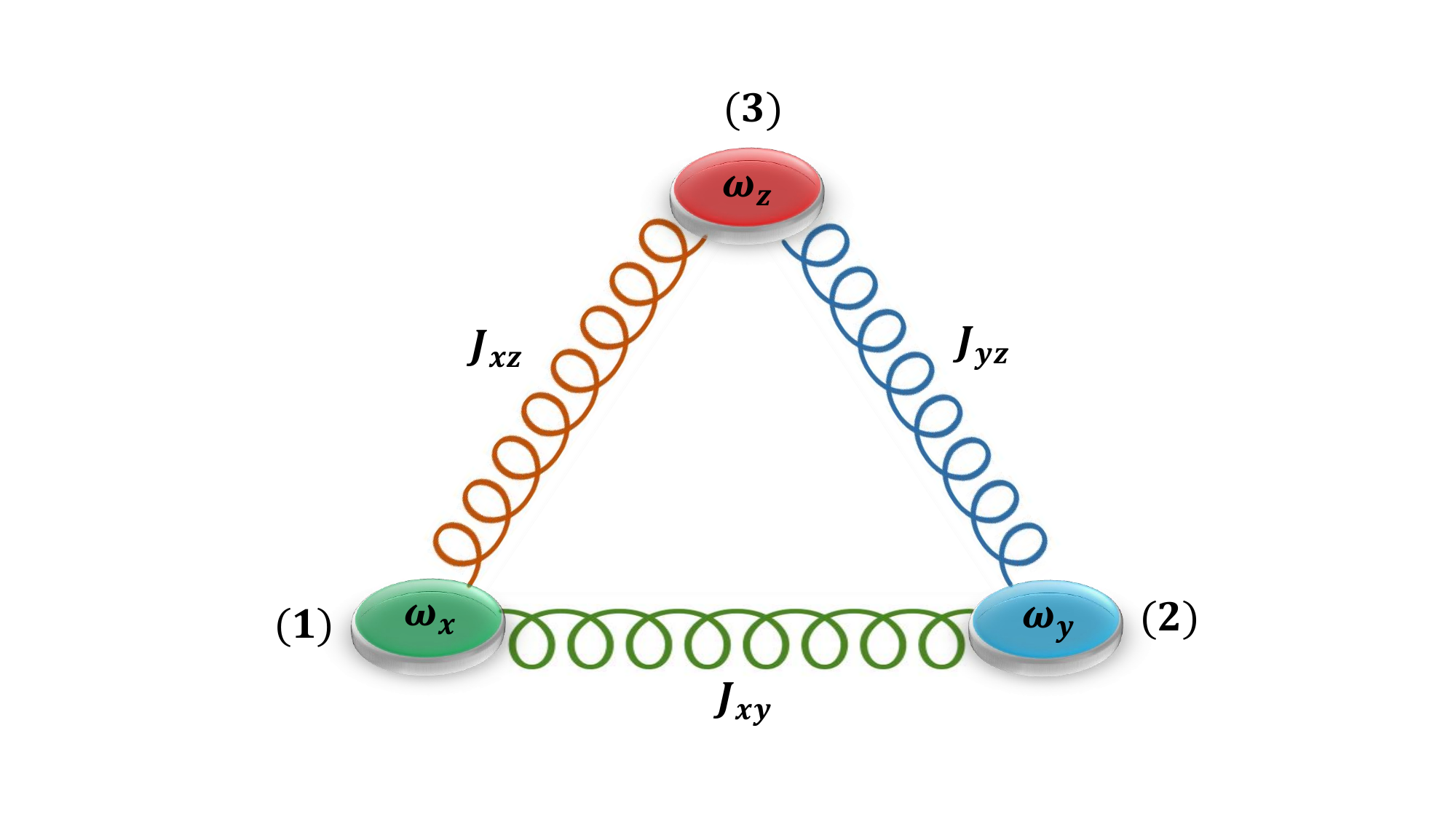}
    \caption{\centering Schematic of three coupled oscillators; each two oscillators are coupled via "position-position" interaction \(\hat{i}\hat{j}\) with a coupling strength of \(J_{ij}\), for all \(i \neq j\) and \(i, j \in \{x, y, z\}\).}
    \label{fig2}
\end{figure}

To address the aforementioned questions, we require a tailored analytical approach. In fact, we begin by decomposing the rotation matrix as: 
\begin{align}
    \mathbb{R}(\varphi, \Phi, \theta) = \mathbb{R}_1(\varphi)\mathbb{R}_2(\Phi)\mathbb{R}_3(\theta)= \left(
\begin{array}{ccc}
1 & 0 & 0 \\
0 & C_\varphi & S_\varphi \\
0 & -S_\varphi & C_\varphi
\end{array}
\right)
\left(
\begin{array}{ccc}
C_\Phi & S_\Phi & 0 \\
- S_\Phi & C_\Phi & 0 \\
0 & 0 & 1
\end{array}
\right)
\left(
\begin{array}{ccc}
C_\theta & 0 & S_\theta \\
0 & 1 & 0 \\
- S_\theta & 0 & C_\theta
\end{array}
\right).\label{eq6}
\end{align}
On one hand, by using the Hamiltonian in Eq.(\ref{eq1}), one can define the so-called potential matrix, namely $\mathbb{K}$. It is defined as: 
\begin{equation}
   \mathbb{K}= \left(
\begin{array}{ccc}
 \omega _x^2 & J_{xy} & J_{xz} \\
 J_{xy} & \omega _y^2 & J_{yz} \\
 J_{xz} & J_{yz} & \omega _z^2 \\
\end{array}
\right).
\end{equation}
On the other hand, the potential matrix can be expressed as $\mathbb{K} = \mathbb{R} \,\Sigma \,\mathbb{R}^{-1}$, where $\Sigma = \mathrm{diag}(\vartheta_x^2, \vartheta_y^2, \vartheta_z^2)$, with $\vartheta_a^2$ representing the eigenvalues of the potential matrix. Additionally, it is worth noting that the explicit expressions of $\vartheta_a^2$ (see \cite{intro8}) are omitted here, as they play no essential role in the subsequent analysis. Then, the diagonal matrix $\Sigma$ can be written as below:
\begin{align}
     \Sigma=\mathbb{R}^{-1}\, \mathbb{K}\, \mathbb{R} &= \mathbb{R}_3(-\theta)\mathbb{R}_2(-\Phi)\mathbb{R}_1(-\varphi)\,\mathbb{K}\,\mathbb{R}_1(\varphi) \mathbb{R}_2(\Phi) \mathbb{R}_3(\theta)\notag\\
     &= \mathbb{R}_3(-\theta)\mathbb{R}_2(-\Phi)\, \mathbb{A}_{\varphi} \, \mathbb{R}_2(\Phi) \mathbb{R}_3(\theta). \label{eq8}
\end{align}
By employing Eq. (\ref{eq6}), one can derive two alternative representations of the auxiliary matrix $\mathbb{A}_{\varphi}$. This step is essential, as it yields such equivalent forms that enable the reduction of the number of independent mixing angles. The auxiliary matrix $\mathbb{A}_{\varphi}$ turns out to be:
\begin{align}
     \mathbb{A}_{\varphi}&=\mathbb{R}_1(-\varphi)\,\mathbb{K}\,\mathbb{R}_1(\varphi)=\mathbb{R}_3(\theta)\mathbb{R}_2(\Phi)\, \Sigma \, \mathbb{R}_2(-\Phi) \mathbb{R}_3(-\theta). \label{eq9}
\end{align}
It is important to note that both $\mathbb{A}_{\varphi}$ and $\mathbb{K}$ are symmetric. Consequently, their inverses coincide with their transposes, i.e., $\mathbb{A}_{\varphi}^{-1} = \mathbb{A}_{\varphi}^T$ and $\mathbb{K}^{-1} = \mathbb{K}^T$, where the superscript $T$ denotes the matrix transpose. The next step is to disentangle the mixing angles. To achieve this goal, we express explicitly the auxiliary matrix $\mathbb{A}_{\varphi}$ according to Eq. (\ref{eq9}) in two distinct representations. The first representation is wrtitten as:
\begin{equation}
   \mathbb{A}_{\varphi}=\left(
\begin{array}{ccc}
\omega_x^2 & J_{xy} C_\varphi - J_{xz} S_\varphi & J_{xy} S_\varphi + J_{xz} C_\varphi \\
J_{xy} C_\varphi - J_{xz} S_\varphi & -2 J_{yz} S_\varphi C_\varphi + C_\varphi^2 \omega_y^2 + S_\varphi^2 \omega_z^2 & J_{yz} C_{2\varphi} + S_\varphi C_\varphi (\omega_y^2 - \omega_z^2) \\
J_{xy} S_\varphi + J_{xz} C_\varphi & J_{yz} C_{2\varphi} + S_\varphi C_\varphi (\omega_y^2 - \omega_z^2) & J_{yz} S_{2\varphi} + S_\varphi^2 \omega_y^2 + C_\varphi^2 \omega_z^2
\end{array}
\right).\label{eq10}
\end{equation}
However, by utilizing the right-hand side of Eq. (\ref{eq9}), the auxiliary matrix $\mathbb{A}_{\varphi}$ can alternatively be expressed as: 
\begin{equation}
  \mathbb{A}_{\varphi}= \left(
\begin{array}{ccc}
C_\theta^2 (C_\Phi^2 \vartheta_x^2 + S_\Phi^2 \vartheta_y^2) + S_\theta^2 \vartheta_z^2 &
C_\theta S_\Phi C_\Phi (\vartheta_y^2 - \vartheta_x^2) &
S_\theta C_\theta (-C_\Phi^2 \vartheta_x^2 - S_\Phi^2 \vartheta_y^2 + \vartheta_z^2) \\
C_\theta S_\Phi C_\Phi (\vartheta_y^2 - \vartheta_x^2) &
S_\Phi^2 \vartheta_x^2 + C_\Phi^2 \vartheta_y^2 &
S_\theta S_\Phi C_\Phi (\vartheta_x^2 - \vartheta_y^2) \\
S_\theta C_\theta (-C_\Phi^2 \vartheta_x^2 - S_\Phi^2 \vartheta_y^2 + \vartheta_z^2) &
S_\theta S_\Phi C_\Phi (\vartheta_x^2 - \vartheta_y^2) &
S_\theta^2 (C_\Phi^2 \vartheta_x^2 + S_\Phi^2 \vartheta_y^2) + C_\theta^2 \vartheta_z^2
\end{array}
\right).\label{eq11}
\end{equation}
Hence, the following identities can be easily demonstrated using both forms of $\mathbb{A}_{\varphi}$: 
\begin{eqnarray}
 \tan(\theta)  &=& -\frac{\mathbb{A}_{23}}{\mathbb{A}_{12}} 
= -\frac{J_{yz}\cos(2\varphi) + \sin(\varphi)\cos(\varphi)(\omega_y^2 - \omega_z^2)}
{J_{xy}\cos(\varphi) - J_{xz}\sin(\varphi)}
\label{eq12},\nonumber\\
\sin(2\Phi) &=& \frac{2\sqrt{\mathbb{A}_{12}^2 + \mathbb{A}_{23}^2}}{\vartheta_x^2 - \vartheta_y^2}
= \frac{2\sqrt{\big(J_{xy}\cos(\varphi) - J_{xz}\sin(\varphi)\big)^2 
		+ \Big(J_{yz}\cos(2\varphi) + \tfrac{1}{2}\sin(2\varphi)(\omega_y^2 - \omega_z^2)\Big)^2}}
{\vartheta_x^2 - \vartheta_y^2}, \nonumber\\
\cos(2\Phi) &=& \frac{\mathbb{A}_{11} - \mathbb{A}_{22} + \mathbb{A}_{33} - \vartheta_z^2}{\vartheta_x^2 - \vartheta_y^2}
= \frac{2 J_{yz}\sin(2\varphi) + \omega_x^2 + \cos(2\varphi)(\omega_z^2 - \omega_y^2) - \vartheta_z^2}
{\vartheta_x^2 - \vartheta_y^2},\nonumber \\
\tan(2\Phi) &=& \frac{2\sqrt{\mathbb{A}_{12}^2 + \mathbb{A}_{23}^2}}
{\mathbb{A}_{11} - \mathbb{A}_{22} + \mathbb{A}_{33} - \vartheta_z^2}
= \frac{2\sqrt{\big(J_{xy}\cos(\varphi) - J_{xz}\sin(\varphi)\big)^2 
		+ \Big(J_{yz}\cos(2\varphi) + \tfrac{1}{2}\sin(2\varphi)(\omega_y^2 - \omega_z^2)\Big)^2}}
{2 J_{yz}\sin(2\varphi) + \omega_x^2 - \cos(2\varphi)\,\omega_y^2 + \cos(2\varphi)\,\omega_z^2 - \vartheta_z^2}.
\label{eq15}
\end{eqnarray}
If the angle $\varphi$ is specified, the corresponding angles $\Phi$ and $\theta$ can be determined from Eqs.~(\ref{eq10}) and (\ref{eq11}).  
Therefore, to obtain an explicit formula for $\varphi$ in terms of the physical parameters $\mathbb{K}_{ij}$ $(i \neq j)$, we make use of Eq.~(\ref{eq8}), such that: 
\begin{equation}
    \Sigma_{ij}=0, \quad (i \neq j).
\end{equation}
After a straightforward sequence of algebraic manipulations (which are too long to be reported here), one can obtain the following condition: 
\begin{equation}
    \mathbb{A}_{13} \left( \mathbb{A}_{12}^2 - \mathbb{A}_{23}^2 \right) 
    = \mathbb{A}_{23} \, \mathbb{A}_{12} \left( \mathbb{A}_{11} - \mathbb{A}_{33} \right).
\end{equation}
This equation leads to 
\begin{equation}
\sum_{n=0}^{5} a_n \mu_{\varphi}^n = 0, \quad \text{where} \quad \mu_{\varphi} = \tan(\varphi),\label{varphi}
\end{equation}
where
\begin{eqnarray}
 a_0&=& J_{xy} J_{yz} \left(\omega _z^2-\omega _x^2\right)+J_{xy}^2 J_{xz}-J_{xz} J_{yz}^2, \nonumber\\
a_1&=& J_{xy} \left(-2 J_{xz}^2+J_{yz}^2+\left(\omega _x^2-\omega _z^2\right) \left(\omega _z^2-\omega _y^2\right)\right)+J_{xz} J_{yz} \left(\omega _x^2-2 \omega _y^2+\omega _z^2\right)+J_{xy}^3, \nonumber\\
a_2&=& J_{xz} \left(\omega _x^2-\omega _y^2\right) \left(\omega _y^2-\omega _z^2\right)-J_{xy}^2 J_{xz}+J_{xy} J_{yz} \left(\omega _y^2-\omega _z^2\right)+J_{xz}^3, \nonumber\\
a_3&=& -J_{xy} \left(J_{xz}^2+\left(\omega _x^2-\omega _z^2\right) \left(\omega _y^2-\omega _z^2\right)\right)+J_{xy}^3+J_{xz} J_{yz} \left(\omega _z^2-\omega _y^2\right), \nonumber\\
a_4&=& J_{xy} J_{yz} \left(\omega _x^2+\omega _y^2-2 \omega _z^2\right)+J_{xz} \left(J_{xz}^2+J_{yz}^2+\left(\omega _x^2-\omega _y^2\right) \left(\omega _y^2-\omega _z^2\right)\right)-2 J_{xy}^2 J_{xz}, \nonumber\\
a_5&=& J_{xz} J_{yz} \left(\omega _y^2-\omega _x^2\right)+J_{xy} \left(J_{xz}^2-J_{yz}^2\right).
\end{eqnarray}
In realistic physical platforms, the couplings between distinct subsystems are generally in the same order, i.e., $J_{xy}\sim J_{xz} \sim J_{yz}$ and much weaker than the characteristic transition frequencies \cite{Maka1,Maka2,Maka3,RAD1}. Then, the couplings and oscillator frequencies satisfy the following assumption: 
\begin{equation}
\max(J_{xy}, J_{xz}, J_{yz}) \ll \min(\omega_x, \omega_y, \omega_z),\label{weak2}
\end{equation}
by using the eigenvalues reported in \cite{intro8}, we lead to the following approximate relations: 
\begin{equation}
\omega_x \sim \omega_y \sim \omega_z \sim \vartheta_x \sim \vartheta_y \sim \vartheta_z\sim\vartheta. \label{weak1}
\end{equation}
Under these physically motivated assumptions, the parameters $a_n$ vanish and Eq.~(\ref{varphi}) is satisfied for any value of $\mu_{\varphi}$, yielding to $\varphi \in \left]-\frac{\pi}{2}, \frac{\pi}{2}\right[$.  Furthermore, by employing Eq.~(\ref{eq12}), we obtain one of the central results of this work, namely explicit expressions for the remaining mixing angles in terms of $\varphi$,  
\begin{eqnarray}
    \tan (\theta )&=& -\sin (\varphi )-\cos (\varphi ), \nonumber\\
    \tan (2 \Phi )&=& \frac{ \sqrt{1-2 \sin (\varphi ) \cos (\varphi )}\sqrt{\sin (2 \varphi )+2}}{\sin (2 \varphi )}.
\end{eqnarray}
Hence, we demonstrate that the behavior of our three coupled quantum oscillators can be fully characterized using a single mixing angle $\theta$. In this framework, it is important to mention that the relationship between $\mu_{\Phi}$, $\mu_{\varphi}$, and $\mu_{\theta}$ is given as: 
\begin{equation}
    \mu _{\Phi }=\frac{1-\mu _{\theta }^2-\sqrt{3-\mu _{\theta }^2}}{\sqrt{2+\mu _{\theta }^2-\mu _{\theta }^4}}, \quad \mu _{\varphi }= \frac{1-\mu _{\theta } \sqrt{2-\mu _{\theta }^2}}{\mu _{\theta }^2-1}, \label{27}
\end{equation}
where $\mu_{\Phi}=\tan(\Phi)$, $\mu_{\theta}=\tan(\theta)$ and $\mu_{\varphi}=\tan(\varphi)$. This significantly simplifies the analysis of entanglement within the system. To go further, the eigenenergies of the system reduce to: 
\begin{equation}
E(n,m,l) = \vartheta_x \left( n + \frac{1}{2} \right) + \vartheta_y \left( m + \frac{1}{2} \right)+\vartheta_z \left( l + \frac{1}{2} \right)\simeq  \vartheta(n+m+l+\frac{3}{2}).
\end{equation}
The corresponding eigenfunctions are expressed as: 
\begin{equation}
\Psi_{(n,m,l)}(X, Y,Z) = \frac{1}{\sqrt{2^{n+m+l} n! m! l!}} \left( \frac{\vartheta}{\pi} \right)^{\frac{3}{4}} 
\exp\left({-\frac{\vartheta}{2} (X^2+Y^2+Z^2)}\right) 
H_n(\sqrt{\vartheta} X) H_m(\sqrt{\vartheta} Y)H_l(\sqrt{\vartheta} Z),
\end{equation}
where the new position coordinates are written as: 
\begin{eqnarray}\label{eq26}
   X &=& x C_\theta C_\Phi + y (-C_\theta C_\varphi S_\Phi - S_\theta S_\varphi) + z (C_\theta S_\varphi S_\Phi - S_\theta C_\varphi), \nonumber \\
   Y &=& x S_\Phi + y C_\varphi C_\Phi - z S_\varphi C_\Phi, \nonumber\\
   Z &=& x S_\theta C_\Phi + y (C_\theta S_\varphi - S_\theta C_\varphi S_\Phi) + z (S_\theta S_\varphi S_\Phi + C_\theta C_\varphi), 
\end{eqnarray}
with their corresponding momenta take the following form: 
\begin{eqnarray}
    P_x &=&  p_x C_\theta C_\Phi +p_y (-C_\theta C_\varphi S_\Phi - S_\theta S_\varphi) + p_z (C_\theta S_\varphi S_\Phi - S_\theta C_\varphi), \nonumber \\
   P_y &=& p_x S_\Phi + p_y C_\varphi C_\Phi-p_z S_\varphi C_\Phi, \nonumber \\
   P_z &=&p_x S_\theta C_\Phi + p_y (C_\theta S_\varphi - S_\theta C_\varphi S_\Phi)+ p_z (S_\theta S_\varphi S_\Phi + C_\theta C_\varphi). \label{eq32}
\end{eqnarray}
In the following section, we shall investigate the quantum entanglement structure in phase space by employing the Wigner distribution function.

\section{Quantum Entanglement and Phase space formalism\label{sec3}}

As is well known, the first phase-space representation of a quantum state was introduced in 1932 by Eugene Wigner \cite{2nd_chapter1}. Indeed, the Wigner function provides a complete description of a quantum state, offering more comprehensive information than can be extracted through conventional quantum approaches \cite{2nd_chapter2,2nd_chapter3}. Owing to this property, it has become a powerful framework for investigating a wide range of quantum phenomena. Moreover, the entanglement phenomenon can be characterized within this framework either by analyzing the negativity of the Wigner function \cite{Ngo,Ngo1} or by evaluating the quantum purity of the reduced (marginal) states \cite{adesso1,adesso2}. For the sake of simplicity, given the specific structure of our quantum state, we adopt the latter approach in this work. To proceed further, the Wigner function associated with our diagonalized Hamiltonian is separable \cite{kim}. Thus, one obtains: 
\begin{equation}
W_{(n,m,l)}(X, P_x; Y, P_y;Z,P_z) = W_n(X, P_x) \times W_m(Y, P_y)\times W_l(Z, P_y), \label{eq21}
\end{equation}
where the reduced Wigner functions are defined as \cite{kim}
\begin{eqnarray}
W_n(Q_i, P_i) &=& \frac{1}{\pi} \int \Psi_n^*(Q_i + \mathbb {Q_i}) \Psi_n(Q_i - \mathbb {Q_i}) \exp\left({2iP_i\mathbb {Q_i}}\right) d\mathbb {Q_i}, \nonumber \\
          &=& \frac{(-1)^n}{\pi} \exp\left({-\frac{1}{\vartheta_i} (\vartheta_i^2 Q_i^2 + P_i^2)}\right)\mathcal{L}_n \left[ \frac{2}{\vartheta_i} (\vartheta_i^2 Q_i^2 + P_i^2) \right].\label{eq22}
\end{eqnarray}
Here \( Q_i \in \{X, Y, Z\} \), \( P_i \in \{P_x, P_y, P_z\} \), \(\vartheta_i \in \{\vartheta_x, \vartheta_y, \vartheta_z\} \) and $\mathcal{L}_n(x)$ introduce Laguerre polynomials \cite{2nd_chapter4}. To move forward, we use the Rodrigues formula for Laguerre polynomials \cite{2nd_chapter4,2nd_chapter5}.
\begin{equation}
\mathcal{L}_n(x) = \frac{1}{n!} \frac{d^n}{du^n} \left( \frac{ \exp\left({-\frac{x u}{1-u}}\right)}{1-u} \right) \Bigg|_{u=0}, \label{eq37}
\end{equation}
to get the Wigner function related to our Hamiltonian as follows: 
\begin{align}
\hspace{-1cm} W_{(n,m,l)}(X, P_x; Y, P_y; Z, P_z) &= \mathcal{R}_{n,l,m} (u,s,v) \left( -\frac{\exp \left(\frac{(v+1) \left(P_z^2 + Z^2 \vartheta_z^2 \right)}{(v-1) \vartheta_z} + \frac{(u+1)(P_x^2+X^2 \vartheta_x^2)}{(u-1) \vartheta_x} + \frac{(P_y^2+Y^2\vartheta_y^2) (s+1)}{(s-1) \vartheta_y}\right)}{(s-1) (u-1) (v-1)} \right)\Bigg|_{u,s,v=0},
\label{eq38}
\end{align}
where the operator $\mathcal{R}_{(n,l,m)}$ is written as : 
\begin{equation}
 \mathcal{R}_{n,l,m} (u,s,v)=\frac{ (-1)^{l+m+n}}{\pi^3 l! m! n!} \frac{d^n}{du^n} \frac{d^m}{ds^m} \frac{d^l}{dv^l}.\label{eq39}
\end{equation}
Under the realistic assumption in Eqs.(\ref{weak1}) and (\ref{weak2}), the Wigner function reads: 
\begin{equation}
W_{(n,m,l)}(X, P_x; Y, P_y; Z, P_z)= \mathcal{R}_{n,l,m} (u,s,v) \left( -\frac{\exp \left(\frac{1}{\vartheta}\left[\frac{(v+1) \left(P_z^2 + Z^2 \vartheta^2 \right)}{(v-1)} + \frac{(u+1)(P_x^2+X^2 \vartheta^2)}{(u-1)} + \frac{(P_y^2+Y^2\vartheta^2) (s+1)}{(s-1)}\right]\right)}{(s-1) (u-1) (v-1)} \right)\Bigg|_{u,s,v=0}.
\end{equation}
Given that the global state is pure, as a consequence of : 
\begin{equation}
    \mathbb{P} = (2\pi)^3 \int_{\mathbb{R}^6} dx\, dp_x\,dy\,dp_y\,dz\,dp_z\, W^2_{(n,m,l)}(x,p_x;y,p_y;z,p_z)=1
\end{equation}

Therefore, the analysis of quantum entanglement in our system, namely the three coupled harmonic oscillators, is carried out by evaluating the purities of the corresponding reduced states 
\cite{adesso1,adesso2,kim}:
\begin{eqnarray}
    \mathbb{P}_x(n,m,l) &=& 2\pi \int_{\mathbb{R}^2} dx\, dp_x\, W^2_{(n,m,l)}(x,p_x), \nonumber \\
    \mathbb{P}_y(n,m,l) &=& 2\pi \int_{\mathbb{R}^2} dy\, dp_y\, W^2_{(n,m,l)}(y,p_y),\nonumber \\
    \mathbb{P}_z(n,m,l) &=& 2\pi \int_{\mathbb{R}^2} dz\, dp_z\, W^2_{(n,m,l)}(z,p_z),\label{45}
\end{eqnarray}
where the marginal Wigner functions $W_{(n,m,l)}(x,p_x), \,W_{(n,m,l)}(y,p_y)$ and $W_{(n,m,l)}(z,p_z)$ are defined as \cite{kim}
\begin{eqnarray}
   W_{(n,m,l)}(x,p_x) &=& \int_{\mathbb{R}^{4}} dy\,dp_y\,dz\,dp_z\,W_{(n,m,l)}(x,p_x;y,p_y;z,p_z), \nonumber\\
   W_{(n,m,l)}(y,p_y) &=& \int_{\mathbb{R}^{4}} dx\,dp_x\,dz\,dp_z\,W_{(n,m,l)}(x,p_x;y,p_y;z,p_z), \nonumber\\
   W_{(n,m,l)}(z,p_z) &=& \int_{\mathbb{R}^{4}} dx\,dp_x\,dy\,dp_y,W_{(n,m,l)}(x,p_x;y,p_y;z,p_z).\label{48}
\end{eqnarray}
Now, by evaluating the integrals in Eq.~(\ref{48}) and substituting them into the marginal purities expressions,  
followed by computing the integrals in Eq.~(\ref{45}), we arrive finally to our key results as: 
\begin{eqnarray}
     \mathbb{P}_x(n,m,l) &=&{(l!m!n!)^{-2}}\frac{d^n}{du^n} \frac{d^m}{ds^m} \frac{d^l}{dv^l}\frac{d^n}{da^n} \frac{d^m}{db^m} \frac{d^l}{dc^l} \bigg( \frac{\left(\mu _{\theta }^2+1\right) \left(\mu _{\Phi }^2+1\right)}{\Omega_1+\Omega_2} \bigg)\Bigg|_{u,s,v,a,b,c=0}, \nonumber\\
       \mathbb{P}_y(n,m,l) &=&{(l!m!n!)^{-2}}\frac{d^n}{du^n} \frac{d^m}{ds^m} \frac{d^l}{dv^l}\frac{d^n}{da^n} \frac{d^m}{db^m} \frac{d^l}{dc^l} \bigg( \frac{-\left(\mu _{\theta }^2+1\right) \left(\mu _{\varphi }^2+1\right) \left(\mu _{\Phi }^2+1\right)}{\Omega_3+\Omega_4+\Omega_5+\Omega_6} \bigg)\Bigg|_{u,s,v,a,b,c=0}, \nonumber\\
        \mathbb{P}_z(n,m,l) &=&{(l!m!n!)^{-2}}\frac{d^n}{du^n} \frac{d^m}{ds^m} \frac{d^l}{dv^l}\frac{d^n}{da^n} \frac{d^m}{db^m} \frac{d^l}{dc^l} \bigg( \frac{-\left(\mu _{\theta }^2+1\right) \left(\mu _{\varphi }^2+1\right) \left(\mu _{\Phi }^2+1\right)}{\Omega_7+\Omega_8+\Omega_9+\Omega_{10}} \bigg)\Bigg|_{u,s,v,a,b,c=0}, 
 \end{eqnarray}
where we define the different  quantities $\Omega_i$ as follows
\begin{eqnarray}
    \Omega_1&=&(c+1) (v+1) \bigg[(a+1) (u+1) (1-b s) \mu _{\Phi }^2+(b+1) (s+1) (1-a u)\bigg], \nonumber\\
    \Omega_2&=&(a+1) (u+1) \mu _{\theta }^2 \bigg[(c+1) (v+1) (1- b s) \mu _{\Phi }^2+(b+1) (s+1) (1-c v)\bigg], \nonumber\\
      \Omega_3&=&(a+1) (b+1) (s+1) (u+1) (c v-1) \left[\mu _{\theta }^2 \mu _{\Phi }^2 \left(\frac{2}{\mu _{\theta }^2-1}-\mu _{\varphi }\right)^2+\mu _{\Phi }^2+1\right], \nonumber\\
    \Omega_4&=&(b+1) (c+1) (s+1) (v+1) (a u-1) \left[\mu _{\Phi }^2\left(\frac{2}{\mu _{\theta }^2-1}-\mu _{\varphi }\right)^2+\mu _{\Phi }^2\mu _{\theta }^2+\mu _{\theta }^2\right], \nonumber\\
    \Omega_5&=&(a+1) (c+1) (u+1) (v+1) (b s-1) \left(\mu _{\theta }^2+1\right) \left[\frac{2}{\mu _{\theta }^2-1}-\mu _{\varphi }\right]^2, \nonumber\\
    \Omega_{6}&=&2 (b+1) (s+1) \mu _{\theta } \mu _{\Phi } \sqrt{\mu _{\Phi }^2+1} \left(\frac{2}{\mu _{\theta }^2-1}-\mu _{\varphi }\right) \bigg[a \bigg(c (u-v)+u v+2 u+1\bigg)-cv(u+2)+u-c-v\bigg], \nonumber\\
     \Omega_7&=&(a+1) (b+1) (s+1) (u+1) (c v-1) \left(\mu _{\theta }^2 \mu _{\varphi }^2 \mu _{\Phi }^2+\mu _{\Phi }^2+1\right), \nonumber\\
    \Omega_8&=&(b+1) (c+1) (s+1) (v+1) (a u-1) \left( \mu _{\Phi }^2\mu _{\varphi }^2+\mu _{\Phi }^2\mu _{\theta }^2+\mu _{\theta }^2\right), \nonumber\\
    \Omega_9&=&(a+1) (c+1) (u+1) (v+1) (b s-1) \left(\mu _{\theta }^2+1\right) \mu _{\varphi }^2, \nonumber\\
    \Omega_{10}&=&-2 (b+1) (s+1) \mu _{\theta } \mu _{\varphi } \mu _{\Phi } \sqrt{\mu _{\Phi }^2+1} \bigg[a \bigg(c (u-v)+u v+2 u+1\bigg)-cv(u+2)+u-c-v\bigg], 
\end{eqnarray}
where $\mu_{\varphi }$ and $\mu_{\phi }$ depend basically on $\mu _{\theta}$ as mentioned in Eq.~(\ref{27}). By applying a recursive differentiation, one can explicitly derive the quantum purities for various states $(n,m,l)$: 
\begin{eqnarray}
\mathbb{P}_x(1,1,0)&=&\frac{2 \mu _{\theta }^2 \left(\mu _{\Phi }^6-\mu _{\Phi }^4+\mu _{\Phi }^2+3\right) \mu _{\Phi }^2+\mu _{\theta }^4 \left(\mu _{\Phi }^2+1\right)^2 \left(\mu _{\Phi }^4+1\right)+\mu _{\Phi }^8-4 \mu _{\Phi }^6+14 \mu _{\Phi }^4-4 \mu _{\Phi }^2+1}{\left(\mu _{\theta }^2+1\right)^2 \left(\mu _{\Phi }^2+1\right)^4}, \nonumber \\
\mathbb{P}_x(1,0,1)&=&\frac{2}{\left(\mu _{\Phi }^2+1\right)^4} \left(-\frac{6 \mu _{\theta }^2 \left(\mu _{\Phi }^2+1\right)}{\left(\mu _{\theta }^2+1\right)^2}+\frac{\left(\mu _{\theta }^4+4 \mu _{\theta }^2+1\right) \left(\mu _{\Phi }^2+1\right)^2}{\left(\mu _{\theta }^2+1\right)^2}+\frac{12 \mu _{\theta }^4}{\left(\mu _{\theta }^2+1\right)^4}-\left(\mu _{\Phi }^2+1\right)^3\right)+1, \nonumber \\
\mathbb{P}_x(0,1,1)&=&\frac{2 \mu _{\theta }^2 \mu _{\Phi }^2 \left(\mu _{\Phi }^6-\mu _{\Phi }^4+\mu _{\Phi }^2+3\right)+\mu _{\theta }^4 \left(\mu _{\Phi }^8-4 \mu _{\Phi }^6+14 \mu _{\Phi }^4-4 \mu _{\Phi }^2+1\right)+\left(\mu _{\Phi }^2+1\right)^2 \left(\mu _{\Phi }^4+1\right)}{\left(\mu _{\theta }^2+1\right)^2 \left(\mu _{\Phi }^2+1\right)^4}. 
\end{eqnarray}
It is worth emphasizing that the above purities depend exclusively on the quantum numbers $n$, $m$, $l$, and the mixing angle $\theta$, since both $\varphi$ and $\Phi$ are themselves functions of $\theta$ as established in Eq.~(\ref{27}). This demonstrates that our technique, \emph{geometrical diagonalization}, proves to be highly effective in reducing the complexity of the entanglement description. By condensing the dependence of the system to a single mixing angle, i.e., $\theta$, it provides a powerful and suitable tool for analyzing bipartite entanglement in our three coupled oscillator systems.  In the following section, we present our numerical illustrations, supported by extensive physical discussions.

\section{ Numerical Results and Discussions\label{sec4}}

The quantification of bipartite quantum entanglement can be achieved through various measures such as purity, linear entropy, von Neumann entropy, and the Schmidt parameter \cite{adesso1,adesso2,Maka1,Maka2}. Among these, linear entropy offers a particularly convenient and computationally efficient tool, especially within the phase-space formalism. In this inspiration, a direct evaluation of the entanglement presented in each subsystem is performed without requiring the full Schmidt decomposition. In fact, the linear entropy corresponding to the marginal purities is given as \cite{adesso2} 
\begin{eqnarray}
    S_{La}(n,m,l)=1-\mathbb{P}_a(n,m,l), \qquad a=x,y,z.
\end{eqnarray}
Furthermore, it is worth noting that the quantification of quantum entanglement in multipartite systems remains one of the most challenging problems in modern physics. Naturally, this difficulty arises from the multiple ways in which different subsystems can become entangled with the rest of the system \cite{class}. In the following analysis, we focus on the three possible bipartitions of the three coupled harmonic oscillators, namely $(x|yz)$, $(y|xz)$ and $(z|xy)$.

\subsection{Quantum entanglement between $x$ oscillator and $yz$ oscillators}

\begin{figure}[H]
    \centering
    \includegraphics[width=0.32\linewidth]{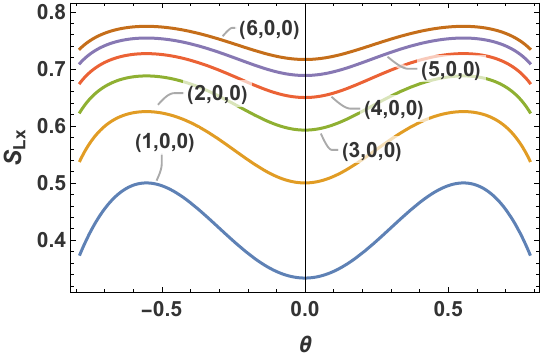}
    \includegraphics[width=0.32\linewidth]{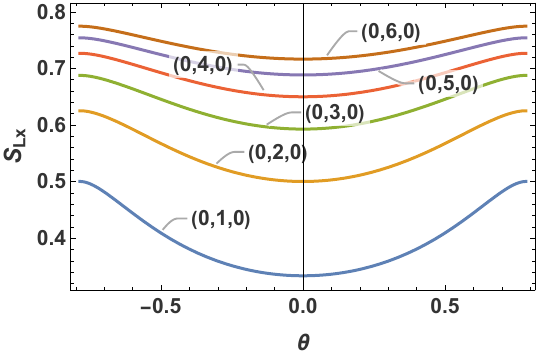}
    \includegraphics[width=0.32\linewidth]{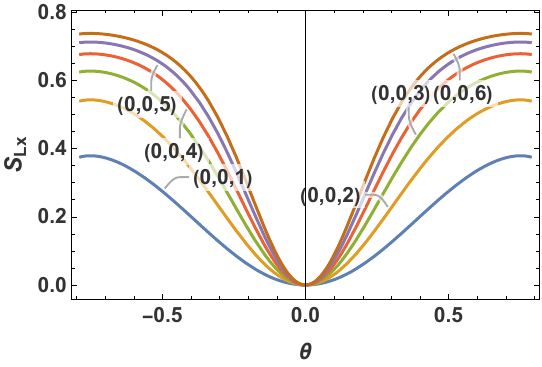}
    \caption{\centering Entanglement $S_{Lx}$ versus $\theta$ for the states $(n,0,0)$, $(0,m,0)$ and $(0,0,l)$ from left to right, respectively.
    \label{figu2} }
\end{figure}

In this part, we shall analyze the entanglement in the bipartition $(x|yz)$, corresponding to the correlations between the oscillator $x$ and the combined subsystem of oscillators $y$ and $z$. Indeed, our investigation begins with the scenario in which the two oscillators are in their ground states while the remaining oscillator is excited. Hence, the marginal linear entropies reduce to: 
\begin{eqnarray}
S_{Lx}[(n,0,0),\theta] &=&1-\frac{(-1)^{(n)}}{16^n n!}(\kappa_1+1)^{2 n} P_n^{(-2 n-1,0)}\left(1-\frac{16 (\kappa_1-1)}{(\kappa_1+1)^2}\right), \nonumber\\
S_{Lx}[(0,m,0),\theta] &=&1-\frac{(-1)^{(m)}}{4^m m!}\left(\frac{\kappa_2-1}{\kappa_2}\right)^{2 m} P_m^{(-2 m-1,0)}\left(\frac{\kappa_2 (\kappa_2+6)+1}{(\kappa_2-1)^2}\right), \nonumber\\
S_{Lx}[(0,0,l),\theta] &=&1-\frac{(-1)^{(l)}}{16^l l!}\left(\frac{\kappa_1 \kappa_2-\kappa_2-2}{\kappa_2}\right)^{2 l} P_l^{(-2 l-1,0)}\left(\frac{16 \kappa_2 (\kappa_1 \kappa_2+\kappa_2-2)}{(-\kappa_1 \kappa_2+\kappa_2+2)^2}+1\right), 
\end{eqnarray}
where $(x)^{(n)}=x(x-1)(x-2)...(x-n+1)$ defines the falling factorial, which is also called lower factorial. Moreover, one can define $\kappa_1$ and $\kappa_2$ as bellows: 
\begin{equation}
  \kappa_1(\theta)=\frac{\left(1-\mu _{\theta }^2\right) \left(\sqrt{3-\mu _{\theta }^2}+2\right)}{\sqrt{3-\mu _{\theta }^2} \left(1+\mu _{\theta }^2\right)}, \quad \kappa_2(\theta)=\frac{\sqrt{3-\mu _{\theta }^2}}{1-\mu _{\theta }^2}.
\end{equation}
Therefore, the amount of entanglement $S_{Lx}$ versus $\theta$  for the states $(n,0,0)$, $(0,m,0)$ and $(0,0,l)$ is displayed in Fig. (\ref{figu2}). It is clear that the quantum entanglement in the bipartition $(x|yz)$ for the state $(n,0,0)$ increases monotonically with respect to $n$, showing that the degree of entanglement grows as the $x$ oscillator becomes more highly excited. For a fixed $n$, the amount of entanglement reaches its maximum at $\mu_\theta = \tfrac{\sqrt{5}-1}{2}$ and its minimum at $\theta=0$.  However, when the $x$ oscillator is prepared in its ground state $(n=0)$, while the $(yz)$ subsystem is in the state $(m\neq0)$, the entanglement exhibits a similar behavior: it attains its maximum for $\mu_\theta=1$ and obtains its minimum at $\mu_\theta=0$.  Finally, when both $x$ and $y$ oscillators remain in their ground states and the $z$ oscillator is excited, the bipartition $(x|yz)$ becomes separable at $\mu_\theta=0$, while reaching maximal entanglement for $\mu_\theta=1$. Physically speaking, these results highlight that the excitations in any of the oscillators amplify the redistribution of correlations within the system. Besides, the parameter $\mu_\theta$ tunes effectively the strength of this entanglement, from complete separability to maximal correlation.

\begin{figure}[H] 
    \centering
    \includegraphics[width=0.32\linewidth]{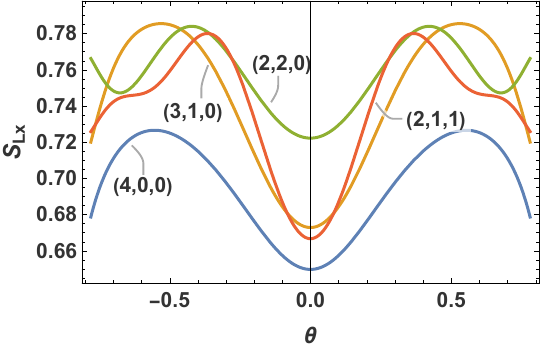}
    \includegraphics[width=0.32\linewidth]{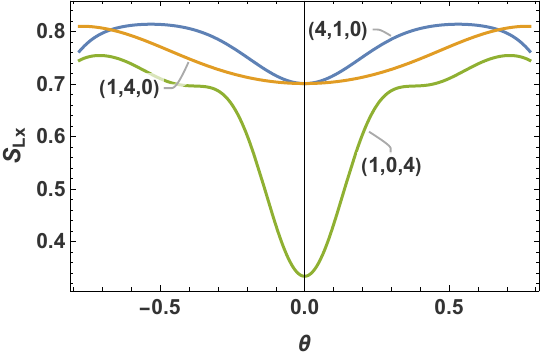}
    \includegraphics[width=0.32\linewidth]{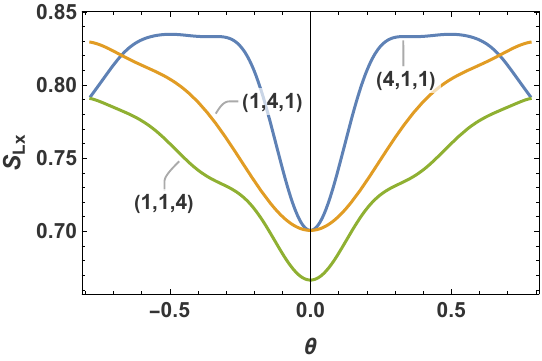}
    \vspace{\baselineskip}
    \includegraphics[width=0.32\linewidth]{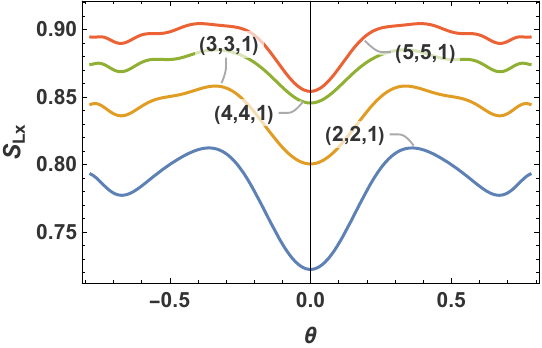}
    \includegraphics[width=0.32\linewidth]{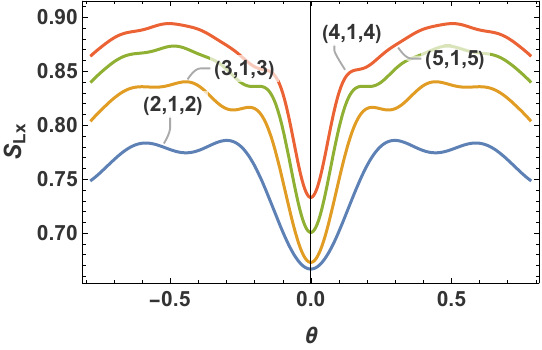}
    \includegraphics[width=0.32\linewidth]{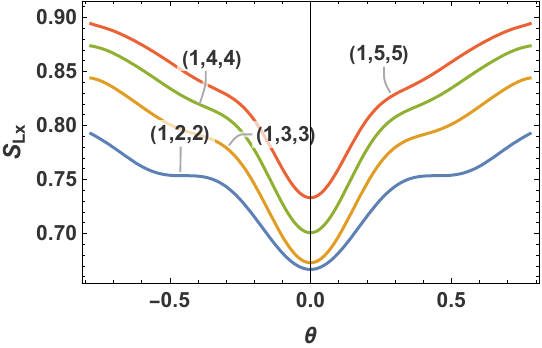}    
    \caption{\centering The evolution of the linear entropy $S_{Lx}$  versus the mixing angle $\theta$, for various quantum  states $(n,m,l)$.
    \label{figure3} }
\end{figure}

In Fig.(\ref{figure3}), we plot entanglement $S_{Lx}$  versus $\theta$ for various states $(n,m,l)$. Obviously, we see that when two oscillators are simultaneously excited, the bipartite entanglement between the subsystem $(x|yz)$ is significantly enhanced compared to the single-excitation case. Moreover, in all configurations, the amount of $S_{Lx}$ increases monotonically with respect to the excitation numbers, reflecting the fact that robust excitations promote stronger quantum correlations. Clearly, the entropy witness reaches its minimum at $\theta=0$, where the state tends toward separability. It attains its maximum at intermediate values of $\theta$, whose precise location depends on the distribution of excitations among the oscillators. This behavior arises physically because simultaneous excitations populate more modes of the coupled system, thereby enriching the structure of quantum correlations and leading to a more robust entanglement profile. In particular, more balanced excitations (e.g., $(2,2,1)$ or $(3,3,1)$) yield smoother and stronger entanglement curves. However, strongly asymmetric configurations still respect the same qualitative features but with a shifted maximum. Another notable feature is that quantum entanglement remains significant over a wide interval of $\theta$ when the $x$ oscillator is highly excited. Note that the state $(4,1,1)$ exhibits a stronger degree of entanglement compared to $(1,4,1)$ and $(1,1,4)$ across most of the $\theta$ range.

From Figs. (\ref{figu2}) and (\ref{figure3}), one can conclude that the amount of entanglement by means of entropy exhibits a pronounced multi-oscillatory behavior, highlighting its high sensitivity to the mixing angle $\theta$. This remarkable tunability suggests that the present system constitutes an excellent candidate for a technological platform that exploits quantum entanglement, offering a natural and versatile hardware implementation.

\subsection{Quantum entanglement between $y$ oscillator and $xz$ oscillators}

Before starting our discussion related to the impact of the parameter $\theta$ on the behavior of entanglement between $y$ oscillator and $xz$ oscillators, let's introduce the exact expressions of the marginal linear entropies. Indeed, they are computed as follows: 
\begin{eqnarray}
	S_{Ly}{[(n,0,0);\theta]}&=&1-\frac{(-1)^{(n)}}{n!}\left(\frac{\kappa_3+\kappa_4}{\kappa_3}\right)^{2 n} P_n^{(-2 n-1,0)}\left(\frac{2 \kappa_4^2}{(\kappa_3+\kappa_4)^2}-1\right), \nonumber\\
	S_{Ly}{[(0,m,0);\theta]}&=&1-\frac{(-1)^{(m)}}{m!}\left(\frac{\kappa_5+\kappa_6}{\kappa_5}\right)^{2 m} P_m^{(-2 m-1,0)}\left(\frac{2 \kappa_6^2}{(\kappa_5+\kappa_6)^2}-1\right), \nonumber\\
	S_{Ly}{[(0,0,l);\theta]}&=&1-\frac{(-1)^{(l)}}{l!}\left(\frac{\kappa_3+\kappa_7}{\kappa_3}\right)^{2 l} P_l^{(-2 l-1,0)}\left(\frac{2 \kappa_7^2}{(\kappa_3+\kappa_7)^2}-1\right), \label{equation70}
\end{eqnarray}
where the quantities $\kappa_i$ reduce to:
\begin{eqnarray}
	\kappa_3(\theta)&=&\left(\mu _{\theta }^2+1\right) \left(\mu _{\Phi }^2+1\right) \left(\left(\frac{2}{\mu _{\theta }^2-1}-\mu _{\varphi }\right)^2+1\right), \nonumber \\
	\kappa_4(\theta)&=& -\left(\frac{2}{\mu _{\theta }^2-1}-\mu _{\varphi }\right)^2 \left(\mu _{\theta }^2 \left(\mu _{\Phi }^2+1\right)+1\right)+2 \left(\frac{2}{\mu _{\theta }^2-1}-\mu _{\varphi }\right) \mu _{\theta }\mu _{\Phi } \sqrt{\mu _{\Phi }^2+1} -\mu _{\Phi }^2-1, \nonumber\\
	\kappa_5 (\theta)&=& \left(\mu _{\Phi }^2+1\right) \left(\left(\frac{2}{\mu _{\theta }^2-1}-\mu _{\varphi }\right)^2+1\right),\nonumber\\
	\kappa_6(\theta)&=&-\left(\frac{2}{\mu _{\theta }^2-1}-\mu _{\varphi }\right)^2\mu _{\Phi }^2 -\mu _{\Phi }^2-1,\nonumber\\
	\kappa_7(\theta)&=&2\left(\frac{2}{\mu _{\theta }^2-1}-\mu _{\varphi }\right) \mu _{\theta } \mu _{\Phi } \sqrt{\mu _{\Phi }^2+1} +\left(\frac{2}{\mu _{\theta }^2-1}-\mu _{\varphi }\right)^2 \left(-\mu _{\theta }^2-\mu _{\Phi }^2-1\right)-\mu _{\theta }^2 \left(\mu _{\Phi }^2-1\right).
\end{eqnarray}

 \begin{figure}[H]
    \centering
    \includegraphics[width=0.32\linewidth]{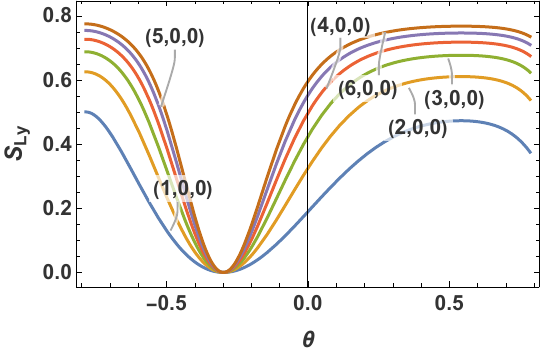}
    \includegraphics[width=0.32\linewidth]{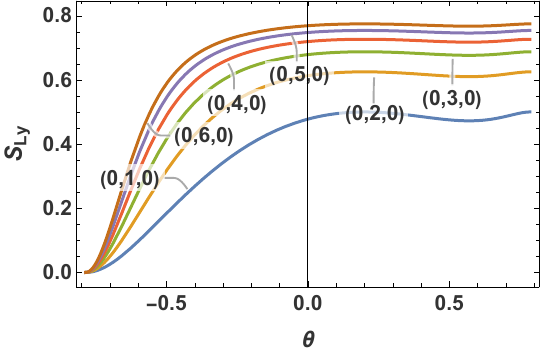}
    \includegraphics[width=0.32\linewidth]{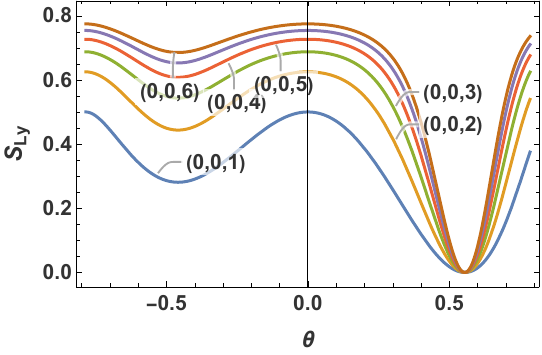}
    \caption{Evolution of the linear entropy $S_{Ly}$ as a function of the mixing angle $\theta$ for the states $(n,0,0)$, $(0,m,0)$, and $(0,0,l)$, shown from left to right, respectively.
    \label{fig3} }
\end{figure}

The amount of entanglement by means of entropy, namely  $S_{Ly}$  versus $\theta$ for the states $(n,0,0)$, $(0,m,0)$ and $(0,0,l)$ is examined in Fig.(\ref{fig3}). As is clear, the entanglement in the bipartition $(y|xz)$ is significant for the state $(n,0,0)$ and becomes important for large quantum numbers $n$, showing that the degree of entanglement grows as the $x$ oscillator becomes more highly excited. However, for a fixed $n$, the entropy vanishes at $\theta \simeq \frac{-2}{21}\pi$. Additionally, for $(n=0)$ and $(m=0)$, the entanglement exhibits a similar behavior, that is, it attains its minimum for $\theta=\frac{11}{62}\pi$. Finally, when both $x$ and $z$ oscillators remain in their ground states and the $y$ oscillator is excited, the bipartition $(x|yz)$ becomes separable at $\theta= -\frac{\pi}{4}$, while keeping increasing over $\theta$. The obtained results demonstrate that excitations in any oscillator enhance the redistribution of correlations throughout the system, while the parameter $\mu_\theta$ modulates efficiently the intensity of this entanglement, ranging from total separability to maximal correlation. Also, it is worth noting that the maximum degree of entanglement is identical for states sharing the same total excitation level, regardless of which oscillator is excited. The most important result here is not the distribution of excitations among the oscillators, but rather the specific values of $\mu_{\theta}$ at which the maximum entanglement is attained.

 \begin{figure}[H]
    \centering
    \includegraphics[width=0.32\linewidth]{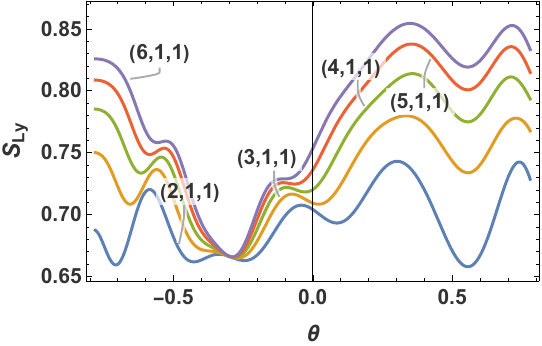}
    \includegraphics[width=0.32\linewidth]{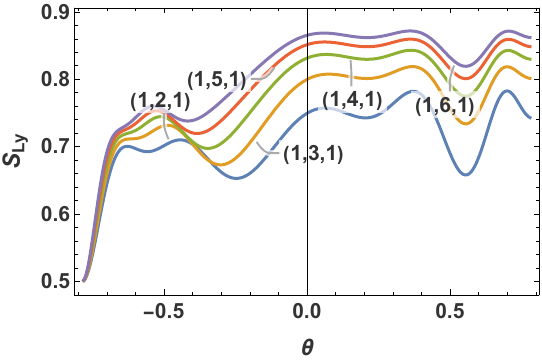}
    \includegraphics[width=0.32\linewidth]{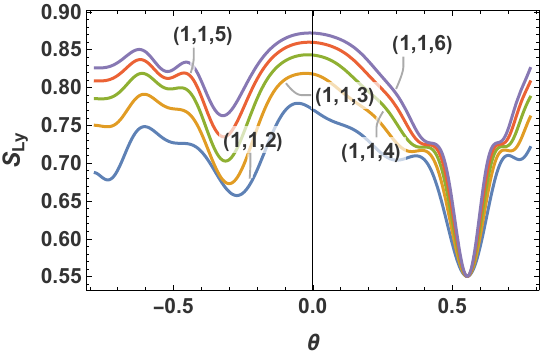}
    \caption{(color online) Evolution of the quantum entanglement $S_{Ly}$ as a function of the mixing angle $\theta$ for the states $(n,1,1)$, $(1,m,1)$, and $(1,1,l)$, shown from left to right, respectively.
 \label{figure5} }
\end{figure}

In Fig.(\ref{figure5}) we evaluate the impact of the parameter $\theta$ on the quantum entanglement $S_{Ly}$ for different states, namely $(n,1,1)$, $(1,m,1)$ and $(1,1,l)$. Obviously, when all oscillators are simultaneously excited, the entropy exhibits a rich multi-oscillatory dependence on the mixing angle $\theta$, a behavior rooted in the intricate structure of the Jacobi polynomials governing the evolution. In this regime, the bipartite entanglement of the $(y|xz)$ subsystem is significantly enhanced compared to the single-excitation case. Across all three panels, increasing excitations systematically increases the entropy, meaning that the excitations generally reinforce entanglement. However, the response to $\theta$ depends strongly on the mode being excited. In the first panel, increasing the first index produces pronounced oscillations with elevated average values. In the second panel, excitations of the second mode enhance entropy but yield more compact non-oscillatory curves. In the third panel, excitations of the third mode induce a sharp dip around $\theta \simeq \tfrac{3\pi}{17}$, where entanglement is considerably suppressed for specific angle values. These results highlight the asymmetric role of different excitation modes, namely, while some invariably strengthen correlations, others introduce strong angle-dependent modulations in the entanglement structure.

\begin{figure}[H]
    \centering
    \includegraphics[width=0.32\linewidth]{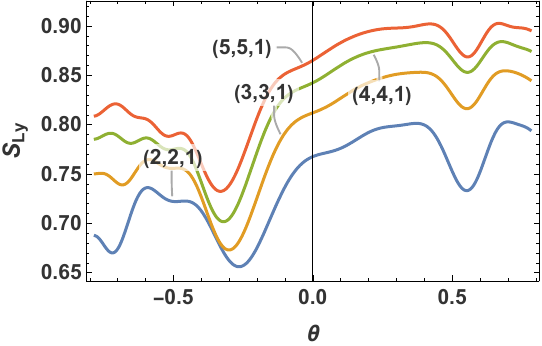}
    \includegraphics[width=0.32\linewidth]{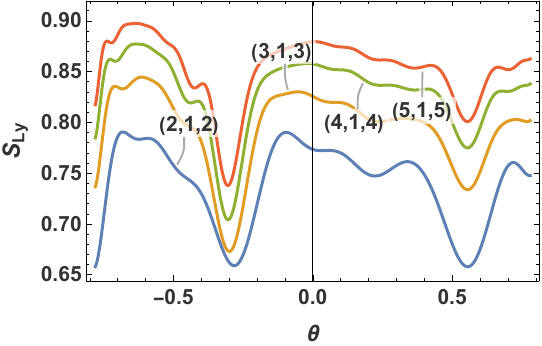}
    \includegraphics[width=0.32\linewidth]{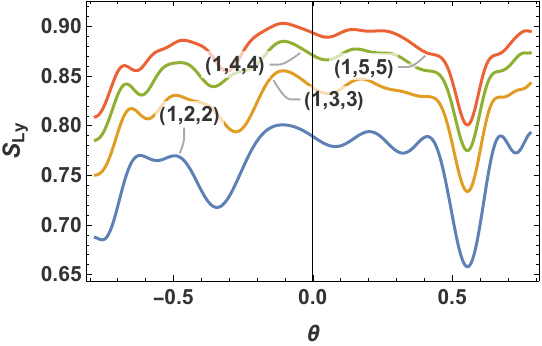}
    \caption{\centering (color online) Evolution of linear entropy $S_{Ly}$ as a function of the mixing angle $\theta$ for the states $(n,m,l)$.
 \label{figure6} }
\end{figure}

The entropy witness $S_{Ly}$  versus $\theta$ but now for the states $(n,m,l)$ is investigated in Fig. (\ref{figure6}). One can observe that there is a clear functional dependence on the mixing angle $\theta$, which confirms that the latter is a controllability parameter of the quantum correlations. The results also reveal an intrinsic asymmetry in the bipartite entanglement $(y|xz)$. It is also apparent that greater excitation states with corresponding greater quantum numbers enhance entanglement, suggesting scaling of quantum correlations with energy in the system. Finally, for a specific value of $\theta$, the entanglement decreases to a minimum value, which is approximately equivalent to the value that is already mentioned in Fig. (\ref{fig3}).

\subsection{Quantum entanglement between $z$ oscillator and $xy$ oscillators}

\begin{figure}[H]
    \centering
    \includegraphics[width=0.32\linewidth]{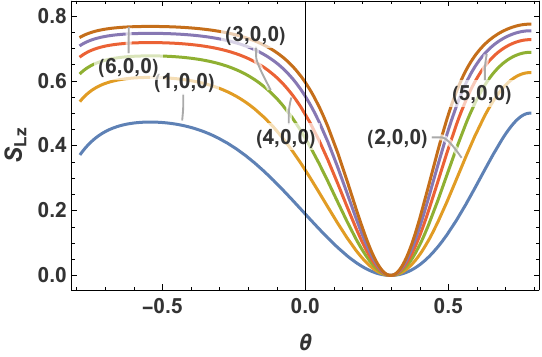}
    \includegraphics[width=0.32\linewidth]{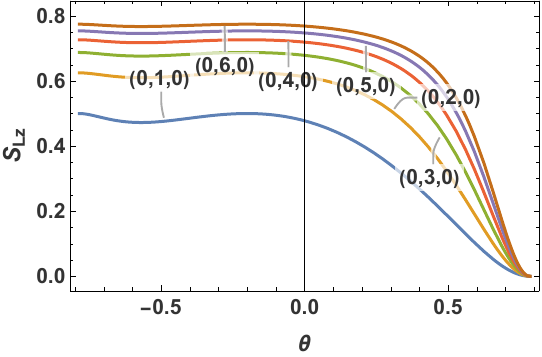}
    \includegraphics[width=0.32\linewidth]{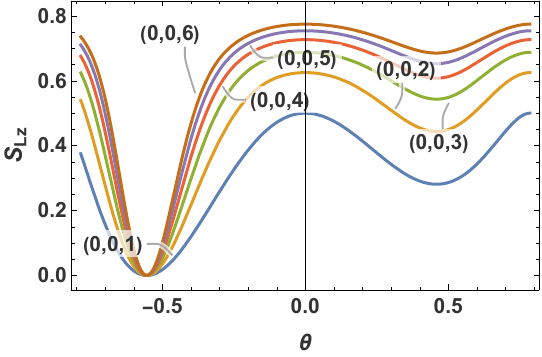}
    \caption{ (color online) Evolution of the linear entropy $S_{Lz}$ as a function of the mixing angle $\theta$ for the states $(n,0,0)$, $(0,m,0)$, and $(0,0,l)$, displayed from left to right, respectively.
    \label{fig4} }
\end{figure}

The results of quantum entanglement computations for the linear entropy $S_{Lz}$ with respect to the bipartition $(xy|z)$ are presented in Fig. (\ref{fig4}). In fact, the bipartite entropies associated with $(xy|z)$ can be expressed as follows: 
\begin{eqnarray}
	S_{Lz}{[(n,0,0);\theta]}&=& S_{Ly}{[(n,0,0);-\theta]}, \nonumber\\
	S_{Lz}{[(0,m,0);\theta]}&=& S_{Ly}{[(0,m,0);-\theta]}, \nonumber\\
	S_{Lz}{[(0,0,l);\theta]}&=& S_{Ly}{[(0,0,l);-\theta]}.
\end{eqnarray}

The results in Fig. ({\ref{fig4}}) show that the evolution of the linear entropy $S_{Lz}$ of the bipartition $(xy|z)$  exhibits a symmetric behavior relative to the linear entropy $S_{Ly}$ for the bipartition $(y|xz)$ for the same states. Furthermore, the quantum entanglement increases monotonically for the state $(n,0,0)$ with $n$. This demonstrates that the entanglement increases as the $x$ oscillator becomes more highly excited. For a fixed value of $n$, the quantum entanglement vanishes at $\theta\simeq \frac{2}{11}\pi$. However, when the $x$ oscillator is given in its ground state, namely $n=0$ but for $(m\neq 0)$, the entanglement increases with higher excited states of $m$. Besides, at a fixed value of $m$, the entanglement decreases with respect to increasing the values of $\theta$ and reaches its minimum at $\theta = \frac{\pi}{4}$. Furthermore, when both the $x$ and $y$ oscillators are in their ground states $(n=0, m=0)$ and the oscillator $z$ is excited $(l\neq 0)$, the entanglement increases with robust values of $l$, where it reaches its minimum at $\theta=-\frac{11}{62}\pi$.

 \begin{figure}[H]
    \centering
    \includegraphics[width=0.32\linewidth]{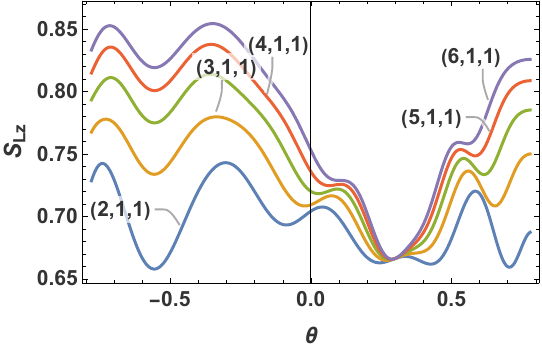}
    \includegraphics[width=0.32\linewidth]{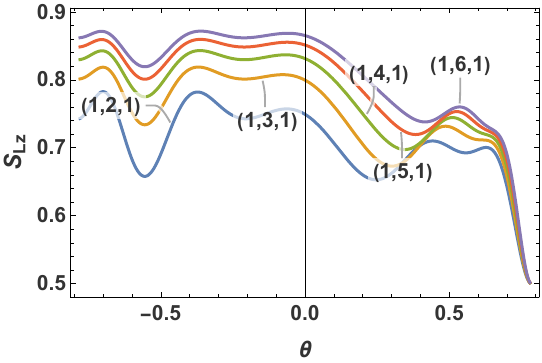}
    \includegraphics[width=0.32\linewidth]{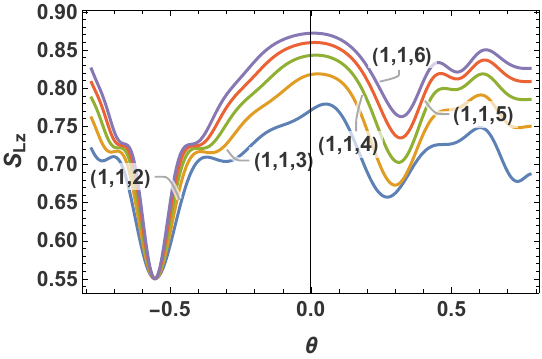}
    \caption{Evolution of the linear entropy $S_{Lz}$ as a function of the mixing angle $\theta$ for the states $(n,1,1)$, $(1,m,1)$, and $(1,1,l)$, displayed from left to right.
  \label{figure8} }
\end{figure} 

 \begin{figure}[H]
    \centering
    \includegraphics[width=0.32\linewidth]{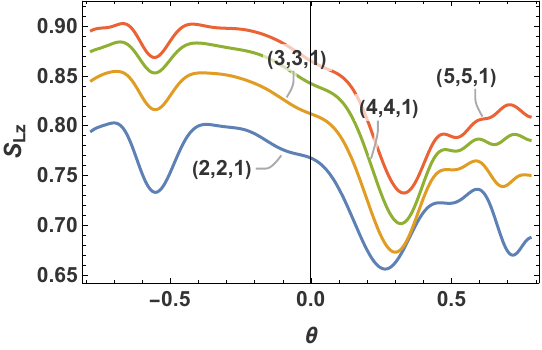}
    \includegraphics[width=0.32\linewidth]{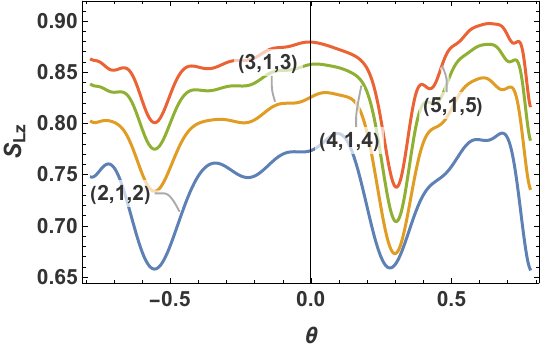}
    \includegraphics[width=0.32\linewidth]{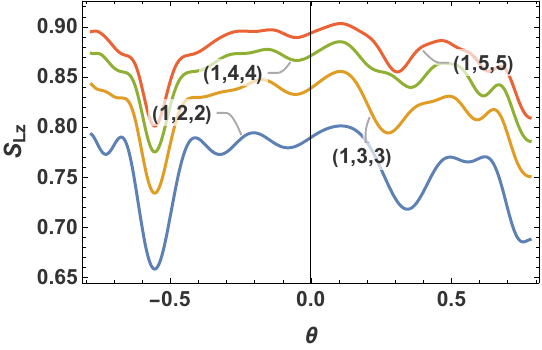}
    \caption{Evolution of linear entropy  $S_{Lz}$  versus the mixing angle $\theta$, for the states $(n,m,l)$.} 
    	\label{figure9}
\end{figure}

In Figs. (\ref{figure8}) and (\ref{figure9}), we study the entropy $S_{Lz}$  versus $\theta$ for the states $(n,1,1)$, $(1,m,1)$, $(1,1,l)$ and $(n,m,l)$, respectively. As displayed, one can observe that the variation of entanglement depends again basically on the mixing angle $\theta$. Furthermore, the results reveal an intrinsic asymmetry in the bipartite entanglement $(xy|z)$. It is also apparent that robust excitation states, with corresponding higher quantum numbers, enhance the entanglement. However, based on the study of the linear entropy of the bipartition $(y|xz)$ and $(xy|z)$, it can be observed that both Figs. (\ref{figure8}) and (\ref{figure9}) display a notable symmetric in their overall structure with respect to that displayed in Figs. (\ref{figure5}) and (\ref{figure6}) for all quantum numbers $n$, $m$ and $l$. Indeed, we can express the relationship between the corresponding linear entropies, namely $S_{Ly}$ and $S_{Lz}$ as: $    S_{Ly}[(n,m,l);\theta]=S_{Lz}[(n,m,l);-\theta]$. This result shows an important symmetry in the quantum information content of the system. In fact, it suggests that the flow of quantum information between these two dual bases is balanced and reversible with respect to the mixing angle $\theta$.

\subsection{Degree of entanglement between one party and the rest}

In this paragraph, we show that quantum entanglement between different bipartitions satisfies a monogamy-like constraint, often referred to as a triangular inequality. Specifically, the amount of entanglement in one bipartition cannot exceed the sum of the entanglements in the other bipartitions. To go further, we introduce the following trade-off quantities \cite{4th_chapter1,4th_chapter2}
\begin{eqnarray}
    M_{k}[(n,m,l),\theta]= S_{Li}[(n,m,l);\theta]+S_{Lj}[(n,m,l);\theta]-S_{Lk}[(n,m,l);\theta], \label{eq83}
\end{eqnarray}
for all distinct $i, j, k \in \{x, y, z\}$.

 \begin{figure}[H]
    \centering
    \includegraphics[width=0.32\linewidth]{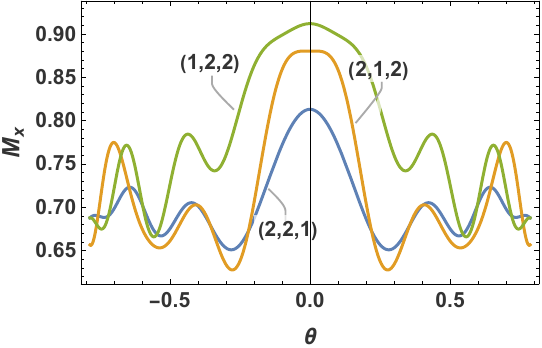}
    \includegraphics[width=0.32\linewidth]{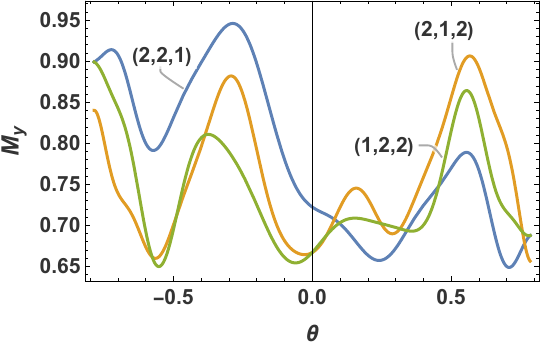}
    \includegraphics[width=0.32\linewidth]{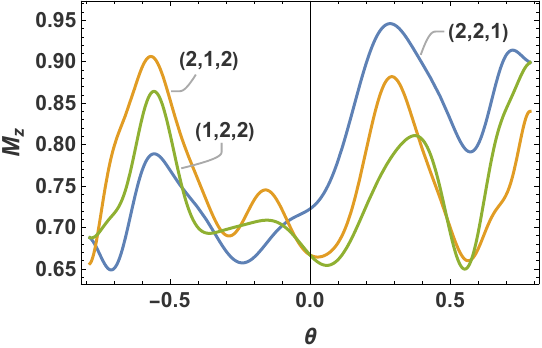}
    \caption{\centering (color online) Evolution of the quantities $M_x$, $M_y$ and $M_z$ versus the mixing angle $\theta$, for the states $(n,m,l)$}
    \label{fig10}
\end{figure}

In Fig. (\ref{fig10}), we present a numerical illustration of the triangular quantity given in Eq.(\ref{eq83}) for some selected states. The results reveal a remarkable intrinsic symmetry. In particular, for the state $(1,2,2)$, the entropy exhibits higher intensity over most of the $\theta$ range compared to the $(2,1,2)$ and $(2,2,1)$ states, where $M_x$ attains its maximum at $\theta = 0$. Moreover, $M_y$ is characterized by a pronounced intensity of the $(2,2,1)$ state for negative values of $\theta$, while the $(2,1,2)$ state dominates for positive values. However, $M_z$ displays an inverse behavior of $M_y$. Then, $M_y$ and $M_z$ exhibit a clear symmetry with respect to one as captured by the relation $M_z(\theta) = M_y(-\theta)$.

\subsection{Two coupled oscillators as a limited case}

Here, let's explore how to derive some results already known in the literature \cite{2nd_chapter6}. Under the conditions where the two angles $\Phi$ and $\varphi$ are both equal to zero, the purity $\mathbb{P}_y$ takes the following form: 
\begin{align}
    \mathbb{P}_y(n,m,l)=1.
\end{align}
Then, the bipartite system $(y|xz)$ is pure, which means that the linear entropy of the $S_{Ly}(n,m,l)$ vanished. The Purities $\mathbb{P}_x$ and $\mathbb{P}_z$ are under the same condition given as :
\begin{align}
 \mathbb{P}(n,l)=\mathbb{P}_x(n,l)= \mathbb{P}_z(n,l)=(n!l!)^{-2}
 \frac{d^n}{du^n}  \frac{d^l}{dv^l}\frac{d^n}{da^n} \frac{d^l}{dc^l}\bigg(-\frac{\mu _{\theta }^2+1}{\Sigma}\bigg)\Bigg|_{u,v,a,c=0}, 
\end{align}
where we define the quantity $\Sigma$ as: 
\begin{equation}
   \Sigma=(c+1) (v+1) (a u-1) \mu _{\theta }^2+(a+1) (u+1) (c v-1). 
\end{equation}
 Moreover, one can also obtain  $\mathbb{P}(n,l)=\mathbb{P}(l,n)$ for all values of $n$ and $l$. This suggests a clear symmetry in the quantum entanglement of the system based on the excitation levels $(n,l)$ of the two oscillators. The results of this analysis are entirely consistent with the analytical formulations for linear entropy and purity defined in Ref.\cite{2nd_chapter6}. In this inspiration, we present the following results of the quantum entanglement computation for the linear entropy for $\Phi=\varphi=0$.

\begin{figure}[H]
    \centering
    \includegraphics[width=0.32\linewidth]{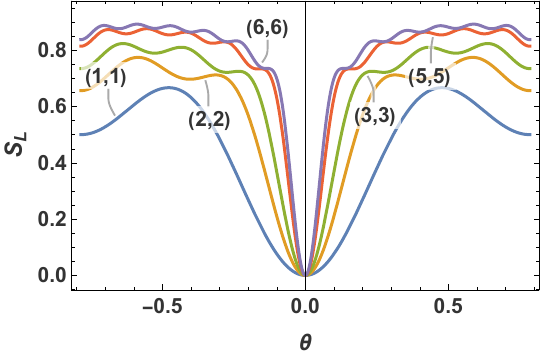}
    \includegraphics[width=0.32\linewidth]{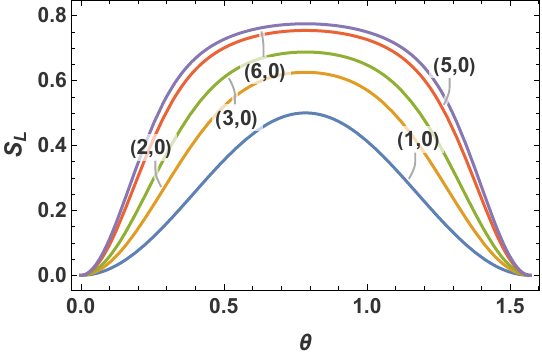}
    \includegraphics[width=0.32\linewidth]{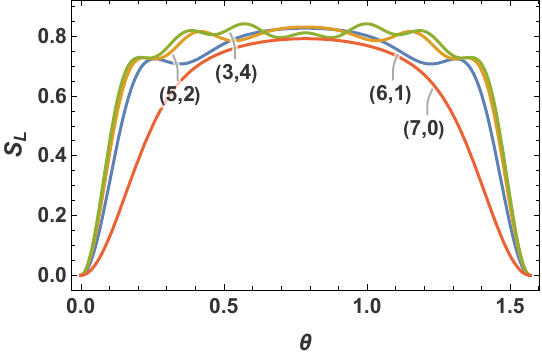}
    \caption{\centering (color online) Evolution of linear entropy versus mixing angle $\theta$ for several bipartite state $(n,m)$.} \label{fig6}
\end{figure} 

Additionally, the purity for the states $(n,0)$ and $(0,n)$ in our case is provided as :
\begin{equation}
	\mathbb{P}(n,0)=\mathbb{P}(0,n)=\frac{(-1)^{(n)}}{n!}\left(\frac{1}{\mu _{\theta }^2+1}\right)^{2 n} P_n^{(-2 n-1,0)}\left(2 \mu _{\theta }^4-1\right).
\end{equation}
By applying a phase shift of $\frac{\pi}{4}$ to the angle to confine it to positive values, we present the corresponding illustration for the same states. 

\begin{figure}[H]
    \centering
    \includegraphics[width=0.32\linewidth]{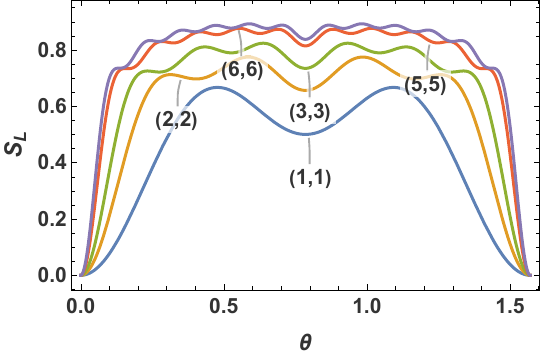}
    \includegraphics[width=0.32\linewidth]{S_L_Enatnglement_of_states_n,0_shifted}
    \includegraphics[width=0.32\linewidth]{S_L_Enatnglement_of_states_n,lwith_n+l=7_shifted}

    \caption{\centering (color online) Evolution of linear entropy $S_{L}$  versus the mixing angle $\theta$, for $\Phi=\varphi=0$, shifted by $\frac{\pi}{4}$.} \label{fig7}
\end{figure} 

Figs. (\ref{fig6}) and (\ref{fig7}) indicate that the initial quantum numbers $n$ and $l$ allow us to enhance basically the entanglement bounds. However, the amount of entanglement does not differ significantly for $n+l=7$. Additionally, for $n=l$, the quantum entanglement achieves its maximum for some values of $\theta$. Also, it is worth noting that this behavior is similar to the one mentioned in \cite{4th_chapter4}.

\section{Conclusion\label{sec5}}

In the present contribution, we proposed a novel mathematical technique termed \emph{geometrical diagonalization}, based on rotational transformation and the mixing angles, in order to investigate the entanglement in three quantum harmonic oscillators. Indeed, we aimed to reduce the degrees of freedom and how this can significantly facilitate quantum entanglement computation and analysis inside this system. Additionally, we used the system's purity to determine the bipartite entanglement of our system based solely on one parameter, namely the mixing angle $\theta$. We demonstrated several particular examples for each bipartite system's entanglement $(x|yz)$, $(y|xz)$, and $(xy|z)$. In addition, we showed that certain outcomes can be expressed in a straightforward analytical manner. We demonstrated that the amount of entanglement for all subsystems is highly sensitive to excitations in any of the oscillators. Indeed, we showed that these excitations enhanced the redistribution of correlations throughout the system.\par

Roughly speaking, the intensity of the entanglement is effectively modulated by the parameter $\theta$, which ranges from maximal correlation to total separability. Furthermore, we exhibited a symmetric behavior between the bipartitions $(y|xz)$ and $(xy|z)$ quantified by means of the equality $S_{Ly}[(n,m,l),\theta]=S_{Lz}[(n,m,l),-\theta]$. Based on this, we obtained a symmetric behavior inside the bipartition $(x|yx)$ for quantum numbers $(n,m,l)$. Moreover, we examined the limiting situation, which involved setting two of the mixing angles ($\Phi$ and $\varphi$) to zero, in order to verify the accuracy of the findings. In conclusion, we have achieved the same entanglement bounds as described in \cite{2nd_chapter6}.\par

Our findings enhanced our understanding of how entanglement behaves in three coupled harmonic oscillators and emphasized the importance of excitation levels and the mixing angle in shaping this quantum phenomenon. These results indicated that quantum oscillators hold promise for quantum communication and quantum technology. The ability to control and manipulate quantum properties may lead to significant advancements in high-precision quantum sensors and computing systems.\\
{\bf{Conflict of Interest}}  \\
The authors declare that they have no conflict of interest.\\

{\bf{Data Availability}} \\
No datasets were generated or analyzed during the current study. Data sharing is not applicable.

\end{document}